\documentstyle[12pt,psfig,epsfig,graphicx,psfrag]{article}
\newcommand{\be}{\begin{equation}}
\newcommand{\ee}{\end{equation}}
\newcommand{\ba}{\begin{eqnarray}}
\newcommand{\ea}{\end{eqnarray}}
\begin{document}

\titlepage
\begin{flushright}
\end{flushright}
\vskip 1cm
\begin{center}
{ \Large \bf Continuous spin and tensionless strings}
\end{center}
\vskip 1cm

\begin{center}
J. Mourad \footnote{mourad@th.u-psud.fr}
\end{center}
\vskip 0.5cm
\begin{center}
{\it Laboratoire
de Physique
Th\'eorique\footnote{Unit\'e Mixte de Recherche du CNRS (UMR 8627).}, B\^at. 210
, Universit\'e
Paris XI, \\ 91405 Orsay Cedex, France\\
and
\\
Laboratoire  Astro Particules et Cosmologie\footnote{Unit\'e Mixte 
de Recherche du CNRS (UMR 7164).}, Universit\'e  Paris VII,\\
2 place Jussieu - 75251 Paris Cedex 05, France.}
\\
\end{center}

\vskip 2cm
\begin{center}
{\large Abstract}
\end{center}
A classical action is proposed which upon quantisation
yields massless particles belonging to the continuous spin
representation of the Poincar\'e group.
The string generalisation of the action is identical to the
tensionless extrinsic curvature action proposed by Savvidy.
We show that the BRST quantisation of the string action
gives a critical dimension of $28$.
The constraints result in  a
number of degrees of freedom which is  the same as  the
bosonic string. 
\noindent

\newpage
\section{Introduction}

Among the irreducible representations of the Poincar\'e group
\cite{Wigner:1939cj},
the massless continuous spin representation
\cite{Wigner:1939cj}-\cite{Brink:2002zx} is very peculiar.
On the one hand, contrary to the other
representations, it does not seem
to describe any physical system, but on the other hand
it seems to be realised by a sector of a string theory
with zero tension. The latter string theory, proposed
by Savvidy  \cite{Savvidy:2003dv,Savvidy:2003fx}, is based on a higher derivative
conformally invariant action with no Nambu-Goto term.

In this article, we first show how to motivate the action
proposed by Savvidy \cite{Savvidy:2003dv,Savvidy:2003fx} starting
 from the action of a point particle
which upon quantisation leads to massless particles belonging to
the continuous spin representation. The point particle action is
proposed in Section 3.
We then analyse in detail the
constraints that arise from the string action.
We show, in Section 4, that the constraints are bilocal.
 They can be transformed
to local ones
for the case of the open string. The resulting spectrum is shown,
in Section 5,
to be given by massless standard helicity states. 
For the case of the closed string, the subject of Section 6, 
we show that the 
bilocal constraints can be transformed to local ones at the
expense of breaking manifest Lorentz invariance. We then 
perform a BRST quantisation \cite{Becchi:1975nq,Henneaux:1985kr}
and show that the nilpotency of the
BRST charge gives a critical dimension of $28$.
The total number of physical degrees of freedom is however $24$,
the same as the transverse coordinates of the bosonic string.
This is a hint of the relevance of this model to describe the
tensionless limit of the bosonic string
\cite{Schild:1976vq}-\cite{Sundborg:2000wp}.
It was conjectured in \cite{Savvidy:2003dv,Savvidy:2003fx}
that the states which belong to the continuous state
representation are all of zero norm. This was verified for the two 
first levels in \cite{Savvidy:2003fx,Antoniadis:2004uh}
using a Gupta-Bleuler covariant quantisation.
in Section 6
we determine the physical states by fixing completely the gauge.
We start in Section 2 by a brief introduction to the
continuous spin representations of the Poincar\'e group.

\section{Fields with a continuous spin representation
of the D-dimensional Poincar\'e group}

Consider a massless  particle in $D$-dimensional spacetime.
Let its momentum have zero components except for $p^+$
($V^{\pm}={1 \over \sqrt{2}}(V^0\pm V^{D-1}),
 V^+=-V_-$). The little group leaving the momentum invariant is
 generated by $M_{ij}$ and $M_{+i}=\pi_i$, they 
 verify the Lie algebra
 \ba
 \left[\pi_i,\pi_j\right] &=& 0,\ \left[M_{ij},M_{kl}\right]=
 \delta_{jk}M_{il}
 -\delta_{ik}M_{jl}-\delta_{jl}M_{ik}+\delta_{il}M_{jk}
 \nonumber\\
 \left[\pi_i,M_{kl}\right]&=&\delta_{ik}\pi_l-\delta_{il}\pi_k,
 \ea
 which is the Lie algebra of the
 $D-2$ dimensional Euclidean group
 $E(D-2)$. The Casimir $\pi^i\pi_i=\mu^2$ classifies the
 representations of $E(D-2)$. If $\mu$ vanishes then the 
 irreducible representations are given by those of $SO(D-2)$,
 these are the helicity states.
 When $\mu$ is nonvanishing we get the
 continuous spin representation. They are of infinite dimension
 and are determined by
 the irreducible representations of the group leaving a given
 $\pi^ i$ invariant, the group $SO(D-3)$.

Wigner \cite{wi,Bargmann:1948ck} proposed manifestly 
covariant wave equations 
 that carry the four-dimensional continuous spin representations.
 The wave function depends, in addition to the usual spacetime
 dependence, on an internal four vector $\xi$.
 The bosonic single valued wave equations are 
 \ba
 p.{\partial \over \partial \xi}\psi=i\mu \psi,\\
 (\xi^2-1)\psi=0,\\
 p.\xi\psi=0,\\
 p^2\psi=0.
 \ea
 Notice that the last two equations can be regarded as
 compatibility conditions for the first two.
 In higher dimensions there are more representations,
 the immediate generalisation of the above equations
  corresponds to
 the trivial representation of $SO(D-3)$. The next one is
 for the vectorial representation of $SO(D-3)$. 
 In order to obtain manifestly covariant equations, one has to
 rely on gauge invariance. One can try
 the equations 
 \ba
 p.{\partial \over \partial \xi}(p_A\psi_B-p_B\psi_A)
 =i\mu (p_A\psi_B-p_B\psi_A),\\
 (\xi^2-1)(p_A\psi_B-p_B\psi_A)=0,\\
 p.\xi(p_A\psi_B-p_B\psi_A)=0,\\
 \xi^A(p_A\psi_B-p_B\psi_A)=0,\ \ 
 p^A(p_A\psi_B-p_B\psi_A)=0.
 \ea
 In fact the first, second and fourth equations are sufficient to
 determine the rest which are compatibility conditions.
To verify that they indeed give  a representation of the 
$SO(D-3)$ group, let us find the solution with the 
spacetime dependence $e^{-ip^+x^-}$. 
By fixing the gauge freedom, we can set $\psi^+=0$.
The third equation gives
$\xi^{-}=0$ and the second $\xi^i\xi_i=1$. The first equation 
gives the dependence on $\xi^+$
\be
\psi_A=e^{-ip^+x^-}\delta(\xi^-)\delta(\xi^i\xi_i-1)
e^{i\mu\xi^{+}/p^+}\phi_A(\hat\xi^i).
\ee
The fourth equation give $\phi^-=0$ and $\xi^i\phi_i=0$.
The solutions of the equations carry the vectorial representation
of the $SO(D-3)$ group.
The generalisation to other representations has to rely on 
gauge invariance and transversality with respect to $\xi$.
A group theoretical treatment can be found in 
\cite{Brink:2002zx}.

\section{Classical action}

We look for a classical action with dynamical variables
$x^\mu$ and $\xi^\mu$ and conjugate momenta $p_\mu$
and $q_\mu$
which gives the constraints
\be
\phi_1=p.q-\mu=0,\ \ \phi_2=\xi^2-1=0,\ \ \phi_3=\xi.p=0,\ \ 
\phi_4=p^2=0.
\ee
Furthermore we require that the action be invariant
under reparametrisation. In order to achieve that we introduce a
one dimensional metric $ds^2=e^2 d\tau^2$ and use it in
the action
\be
S[x^\mu,\xi,e,\lambda]=\int d\tau {\dot x.\dot \xi \over e}
+e\mu +\lambda(\xi^2-1),\label{ac}
\ee
here $\lambda$ is a Lagrange multiplier which gives the 
constraint $\phi_2$.
The momenta associated to $x$ and $\xi$ are
\be
p_\mu={\dot \xi_\mu \over e},\ q_\mu={\dot x_\mu \over e}.
\ee
Variation with respect to $e$ gives
\be
-{\dot x.\dot \xi \over e^2}+\mu=0,
\ee
this is equivalent to $\phi_1$.
The canonical Hamiltonian is 
\be
H_c=p.\dot x+q.\dot \xi+\pi_e\dot e+\pi_\lambda\dot\lambda
-L=e(p.q-\mu)-\lambda(\xi^2-1),
\ee
The condition that the two constraints be preserved in time 
imply that the Poisson brackets of $H_c$ with
$\phi_1$ and $\phi_2$ must vanish. This gives a secondary
constraint \cite{Dirac:1950pj}
$\xi.p=0$ whose preservation in time produces the 
 tertiary constraint  $p^2=0$.
The Poisson bracket of $\phi_1$ and $\phi_2$ gives $\phi_3$
and the Poisson bracket of $\phi_3$ with $\phi_1$ gives $\phi_4$.
It is possible to add Lagrange multiplier to the action in order
to get all the constraints as primary ones
\be
S'[x,\xi,e,\lambda,v_1,v_2]=\int d\tau {\dot x.\dot \xi \over e}
+e\mu +\lambda(\xi^2-1)+v_1{\dot \xi}^2+v_2\xi.\dot\xi.
\ee

An equivalent action can be obtained by eliminating $e$ from the
action (\ref{ac})
to get
\be
S[x,\xi,\lambda]=2\sqrt{\mu}\int d\tau \sqrt{\dot x.\dot \xi} 
+\lambda(\xi^2-1).
\ee
This action has however a singular $\mu\rightarrow 0$ limit.
It is also possible to eliminate the Lagrange multiplier
$\lambda$ by
solving for $\xi^0$ as $\sqrt{\vec \xi^2-1}$ and replacing it in the
first part of the action.
The action loses however its manifest covariance.

Notice that by fixing the gauge $\dot x.\dot\xi=1$
the equations of motion obtained from the action,
after elimination of the variable $\xi$ give a fourth order
equation for $x$ which reads
\be
{d^2 \over d\tau^2}\left[{\ddot x^\mu \over 
\sqrt{\ddot x^\nu \ddot
x_\nu}}\right]=0,
\ee
this equation is also obtained from 
the higher derivative action proposed in \cite{Zoller:1991hs} 
$S=\int d\tau \sqrt{\ddot x^\nu \ddot
x_\nu}$, which is not invariant under reparametrisation\footnote{
Higher derivative actions with reparametrisation invariance
describing fixed helicity massless states
were proposed and studied in \cite{Zoller:1990jv}.}.

\section{String action and resulting constraints}
The world sheet action is now
\be
S[X^\mu,\Xi^\mu,h_{mn},\lambda]=
-\tilde \mu \int d^2\sigma \sqrt{-h}\left[h^{mn}\partial_m X^\mu
\partial_n \Xi^\nu \eta_{\mu\nu} +\lambda(\Xi^2-1)\right],
\label{sac}
\ee
where the two-dimensional metric $h_{mn}$ replaces $e$ and 
the two coordinates $X$ and
$\Xi$ depend on the two world-sheet coordinates $\sigma^0$ and
$\sigma^1$. Notice that we did not add an analog of the second
term in (\ref{ac}) to maintain conformal invariance.
The equations of motion are
\ba
\Box \Xi^\mu=0,\ \Box X^\mu={2\lambda }\Xi^\mu,\label{eqm}
\ea
where $\Box={1 \over
{\sqrt{-h}}}\partial_m\sqrt{-h}h^{mn}\partial_n$.
The constraints obtained by varying with respect to $\lambda$
and $h_{mn}$ are
\ba
(\Xi^2-1)=0,\label{l1}\\ 
\partial_m X^\mu\partial_n\Xi_\mu+\partial_n X^\mu\partial_m \Xi_\mu
-h_{mn}\partial_l X^\mu\partial^l\Xi_\mu=0.
\ea

Using (\ref{eqm}) and (\ref{l1}) we get for $\lambda$
\be
(2\lambda)^2=(\Box X)^2,\label{l2}
\ee

The second equation in (\ref{eqm}) together with
(\ref{l2}) can be used to determine $\Xi$ in terms of $X$ as
\be
\Xi={\Box X \over \sqrt{(\Box X)^2}}.
\ee 
We can eliminate $\Xi$ from the action to get the
higher derivative action
\ba
S&=& -\tilde \mu
\int d^2\sigma \sqrt{-h}h^{mn}\partial_m X^\mu
\partial_n {\Box X_\mu \over \sqrt{(\Box X)^2}}\nonumber\\
&=&\tilde \mu \int d^2\sigma \sqrt{-h}
\sqrt{\Box X^\mu \Box X_\mu},
\ea
which is the form proposed in \cite{Savvidy:2003dv}.

Notice that the kinetic term in the action (\ref{sac})
is invariant under the $SO(D,D)$ inhomogeneous group.
The $(D,D)$ signature of the kinetic terms signals potential
pathologies in the quantum theory like negative norm states and
instabilities. We shall return to this point later. The
constraints, as expected, will be crucial 
in solving this potential difficulty.

We now turn  to determine  the full set of constraints 
contained in the
action. 
The canonical momenta $P_\mu$ and $Q_\mu$ associated respectively
to $X^ \mu$ and $\Xi^\mu$ are (we use the notation
$\gamma^{mn}=\tilde\mu \sqrt{-h}h^{mn}$, we will set from now on
$\tilde \mu=1$)
\be
P_\mu=-\gamma^{00}\partial_0\Xi_\mu-\gamma^{01}\partial_1\Xi_\mu,\ 
Q_\mu=-\gamma^{00}\partial_0 X_\mu-\gamma^{01}\partial_1 X_\mu.
\ee
They allow the determination of the canonical Hamiltonian density
\be
H_c=N(P.Q+\partial_1X.\partial_1\Xi)+
M(P.\partial_1X+
Q.\partial_1\Xi)-\lambda(\Xi^2-1),
\ee
where we have denoted $-(\gamma^{00})^{-1}$ by $N$
and  $\gamma^{01}$ by $M/N$.
The primary constraints are thus
\ba
{\cal H}_0&=&P.Q+\partial_1X.\partial_1\Xi=0\nonumber\\
{\cal H}_1&=&P.\partial_1X+
Q.\partial_1\Xi=0,\nonumber\\
\phi_1&=&\Xi^2-1=0\label{pri}.
\ea
The first two are the usual (Virasoro) constraints due to the
reparametrisation invariance of the string world-sheet. They are
first class. 
The Poisson bracket of ${\cal H}_1$ with $\phi_1$
gives
\be
\{{\cal H}_1(\sigma),\phi_1(\sigma')\}=
\phi_1'(\sigma)\delta(\sigma-\sigma'),
\ee
which is not a new constraint. 
The Poisson bracket of ${\cal H}_0$ with $\phi_1$, on the other
hand,
gives the secondary constraint 
\be
\{{\cal H}_0(\sigma),\phi_1(\sigma')\}=
P.\Xi(\sigma)\delta(\sigma-\sigma')=\phi_2\delta(\sigma-\sigma').
\ee
The new constraint $\phi_2$ does not commute with ${\cal H}_0$,
it generates a tertiary constraint $\phi_3$ which is given by
\be
\phi_3=P^2-(\partial_1\Xi)^2.
\ee
It is easy to convince onself that the procedure does not end.
It is however possible to find all the constraints
by solving explicitly the $\Xi$ equations of motion.
It will be convenient to work in the conformal gauge $M=0, N=1$,
where the solution for $\Xi$ is
\ba
\Xi(\sigma,\tau)&=&\Xi_+(\tau+\sigma)+\Xi_-(\tau-\sigma)\nonumber\\
&=&\Xi(\sigma)+\sum_{n=1}^{\infty}{\tau^n\over{n!}}
\left({(1-(-1)^n)\over 2}P^{(n-1)}(\sigma)
+{(1+(-1)^n)\over 2}\Xi^{(n)}(\sigma)\right)\nonumber\\
&=& {\Xi(\sigma+\tau)+\Xi(\sigma-\tau) \over 2}
+{1 \over 2}\int_{\sigma-\tau}^{\sigma+\tau} d\tilde\sigma\ 
P(\tilde\sigma),
\ea
where $\Xi(\sigma)=\Xi(\sigma,0)$.
From the requirement that $\Xi^2=1$ be true for all time 
we get 
the two bilocal constraints
\ba
(\Xi(\sigma)+\Xi(\sigma')).\int_{\sigma'}^{\sigma}
d\tilde\sigma\ 
P(\tilde\sigma)&=&0,\label{cons1}\\
\Xi(\sigma).\Xi(\sigma')+{1 \over 2}\left(\int_{\sigma'}^{\sigma}
d\tilde\sigma\ 
P(\tilde\sigma)\right)^2=1,\label{cons2}
\ea
where $\sigma$ and $\sigma'$ are arbitrary.
Taking the derivative of (\ref{cons1})
with respect to $\sigma$ gives the equivalent constraint 
\be
\partial_1\Xi(\sigma).\int_{\sigma'}^{\sigma}
d\tilde\sigma\ 
P(\tilde\sigma)+(\Xi(\sigma)+\Xi(\sigma')).P(\sigma)=0.\label{cond}
\ee
If we set $\sigma=\sigma'$ in the above equation we get
\be
P(\sigma).\Xi(\sigma)=0,\label{fi1}
\ee
which is our previous secondary constraint $\phi_2$.
Taking the derivative of (\ref{cond}) with respect to
$\sigma'$ gives
\be
-\partial_1\Xi(\sigma).P(\sigma')+\partial_1\Xi(\sigma').P(\sigma)
=0.\label{bi1}
\ee
Finally we get that  two constraints (\ref{fi1}) and (\ref{bi1})
are equivalent to (\ref{cons1}).

We turn to the second bilocal constraint
(\ref{cons2}). Setting $\sigma=\sigma'$ gives the primary
constraint $\phi_1$.
Its derivative with respect to $\sigma$ gives
\be
\partial_1 \Xi(\sigma).\Xi(\sigma')+P(\sigma).\int_{\sigma'}^{\sigma}
d\tilde\sigma\ 
P(\tilde\sigma)=0. \label{const2}
\ee
Setting $\sigma=\sigma'$ gives the derivative of $\phi_1$
with respect to $\sigma$ and taking the derivative with respect
to $\sigma'$ yields
\be
\partial_1 \Xi(\sigma).\partial_1
\Xi(\sigma')-P(\sigma).P(\sigma')=0.\label{bi2}
\ee
The $\sigma=\sigma'$ part of (\ref{bi2})
gives $\phi_3$. In summary, 
all the constraints are contained in (\ref{pri}),
(\ref{fi1}), (\ref{bi1}) and (\ref{bi2}).
If we expand (\ref{bi1}) and (\ref{bi2}) around $\sigma=\sigma'$
we get an infinite number of local constraints.

An equivalent way of writing the two bilocal constraints
(\ref{bi1}) and (\ref{bi2}) is obtained by taking their sum and
difference.
The sum gives
\be
\chi(\sigma,\sigma')=(P(\sigma)+\partial_1\Xi(\sigma)).(P(\sigma')-
\partial_1\Xi(\sigma'))=0.\label{sum}
\ee
The difference gives the
 equivalent constraint
\be
(P(\sigma)-\partial_1\Xi(\sigma)).(P(\sigma')+
\partial_1\Xi(\sigma'))=0,
\ee
which can be obtained from (\ref{sum}) by
permuting $\sigma$ and $\sigma'$. So both (\ref{bi1}) and
(\ref{bi2}) are contained in (\ref{sum}).

The Poisson brackets of ${\cal H}_0$ 
and ${\cal H}_1$ with $\chi$ can be readily
calculated
\ba
\{{\cal H}_0(\sigma),\chi(\sigma',\sigma'')\}&=&
-\delta'(\sigma-\sigma')\chi(\sigma,\sigma'')+
\delta'(\sigma-\sigma'')\chi(\sigma',\sigma),\\
\{{\cal H}_1(\sigma),\chi(\sigma',\sigma'')\}
&=&-\delta'(\sigma-\sigma')\chi(\sigma,\sigma'')
-\delta'(\sigma-\sigma'')\chi(\sigma',\sigma).
\ea

If we define ${\cal H}_\pm$ as ${\cal H}_{0}\pm{\cal H}_1$,
$P_R(\sigma)=P+\partial_1\Xi,\ P_L(\sigma)=P-\partial_1\Xi$,
then we get
\ba
&&\chi(\sigma,\sigma')=P_R(\sigma).P_L(\sigma'),\
{\cal H}_+=P_R(\sigma).Q_R(\sigma),\ 
{\cal H}_-=P_L(\sigma).Q_L(\sigma),\\
&&\{{\cal H}_+(\sigma),\chi(\sigma',\sigma'')\}=
-2\delta'(\sigma-\sigma')\chi(\sigma,\sigma''), \\
&&\{{\cal H}_-(\sigma),\chi(\sigma',\sigma'')\}=
2\delta'(\sigma-\sigma'')\chi(\sigma',\sigma).
\ea

\section{Open strings}

For open strings with Neumann boundary conditions we have
$\partial_1\Xi(0)=\partial_1\Xi(\pi)=
\partial_1X(0)=\partial_1 X(\pi)=0$.
As usual this implies that the left movers and right movers are
not independent, rather one has
$P_R(\sigma)=P_L(-\sigma)$ and $Q_R(\sigma)=Q_L(-\sigma)$.
The constraints now, keeping the right movers, become
\ba
P_R(\sigma).Q_R(\sigma)&=&0,\quad \Xi^2(\sigma)=1,\\
P_R(\sigma).\Xi(\sigma)&=&P_R(-\sigma).\Xi (\sigma)=0,\quad
P_R(\sigma).P_R(\sigma')=0.
\ea
All the variables are $2\pi$ periodic.

Let us first analyse the last bilocal constraint.
Setting $\sigma=\sigma'$ we get
\be
P_R^-(\sigma)={1 \over 2 P_R^+(\sigma)}P_R^i(\sigma)P_{R\
i}(\sigma).
\ee
Plugging this in the constraint yields
\be
\left(P_R^i(\sigma)-{P^+_R(\sigma)\over P^+(\sigma')}
P_R^i(\sigma')\right)\left(P_{R\ i}(\sigma)-
{P^+_R(\sigma)\over P^+(\sigma')}
P_{R\ i}(\sigma')\right)=0,
\ee
which is verified provided
\be
{P_R^i(\sigma)\over P^+_R(\sigma)}=
{P_R^i(\sigma')\over P^+_R(\sigma')}={p^i\over p^+},
\ee
where $p^\mu$ does not depend  on $\sigma$.
This is equivalent to
\be
P_R^\mu(\sigma)=p^\mu F_R(\sigma),
\ee
that is all the components of $P_R$ are determined by a single
function $F_R$ and a constant momentum $p^\mu$ with $p^2=0$.
The remaining constraints, supposing that $F_R$ does not vanish,
become
\be
p.Q_R=0,\ \  \Xi^{2}-1=0,\ \  p.\Xi=0.\label{op}
\ee 
The function $F_R(\sigma)$ is arbitrary and by
using the conformal invariance can be set to $1$ so that
\be
P^\mu(\sigma)=p^\mu,\quad \Xi^\mu(\sigma)=\xi^\mu,
\ee
where $\xi^\mu$ does not depend on $\sigma$.
The first equation removes the oscillator modes of $X$ as
dynamical variables and the second removes the oscillator modes of
$\Xi$.
The problem has been greatly simplified since
the dynamical variables now are just the zero modes $x$ and $p$ as
well as $\xi$ and its conjugate momentum $q$
 subject to the 
constraints (\ref{op}) which become

\be
p.q=0,\ \xi^2=1,\ p.\xi =0,\ p^2=0.
\ee
where $q$ and $p$ are canonically conjugate to $\xi$ and
$x$.
The quantisation of the system is straightforward:
the wave function $\psi(\xi,p)$ is subject 
to the constraints
\be
(\xi^2-1)\psi=0,\ p.{\partial \over \partial \xi}\psi=0,\
p^2\psi=0.
\ee
The solution can be written as
\be
\psi=\sum_n h_{\mu_1\dots\mu_n}(p)\xi^{\mu_1}\dots\xi^{\mu_n},
\ee
where $h_{\mu_1\dots\mu_n}$ is a traceless
completely symmetric tensor
which verifies the massless spin $n$ equations of motion
(see for instance \cite{Bouatta:2004kk} and references therein)
\be
p^2h_{\mu_1\dots\mu_n}=0,\ \ p^{\mu_1}h_{\mu_1\dots\mu_n}=0.
\ee
Notice that the first constraint is equivalent to the
tracelessness of $h$.
The open string does not possess physical modes with the continuous spin
representation. It is important to note that 
the quantisation reduces to the  point particle 
quantisation and so does not ask for a critical
dimension.

\section{Closed strings}

We start by analysing the bilocal constraint
$\chi(\sigma,\sigma')=P_R(\sigma).P_L(\sigma')=0$.
Let $V_R$ be the vector space spanned by $P_R(\sigma)$ when
$\sigma$ varies from $0$ to $2\pi$ and
similarly for $V_L$, then $V_R$ and $V_L$ are orthogonal.
Let $p$ be the common zero mode of $P_R$ and $P_L$.
By taking the integral on both $\sigma$ and $\sigma'$ of the
constraint $\chi$ we
get that $p^2=0$. All the string modes are thus massless.
Furthermore $p$ is contained in both $V_R$ and
$V_L$. 
Suppose that the only nonvanishing component of $p$ is $p^+$,
and split the spacelike and transverse indices  $i=1,\dots D-2$
into $a$ which belong to $V_R$ and $a'$ which belong to $V_L$.
We thus have
\be
P_L^{a}(\sigma)=0,\ a=1,\dots N,\quad
P_R^{a'}(\sigma)=0,\ a'=N+1,\dots D-2,\label{contl}
\ee
and since $p^+$ is in both $V_R$ and $V_L$ we also have
\be
P_L^{-}(\sigma)=0=P_R^{-}(\sigma).
\ee
The latter two constraints are equivalent to $\Xi^-=\xi^-$
and $P^-=0$, with $\xi^-$ a constant zero mode.
We have transformed the bilocal constraints into local ones at 
the expense of breaking the manifest Lorentz invariance.
Now the constraint $P.\Xi=0$ reads
\be
-P^{+}\xi^-+P^a\Xi_a+P^{a'}\Xi_{a'}=0,\label{pr}
\ee
which gives
\be
-P^{+}\xi^- +\partial_1\Xi^a\Xi_a
-\partial_1\Xi^{a'}\Xi_{a'}=0,
\ee
upon using (\ref{contl}).
Taking the integral over $\sigma$ and using the fact
that $p^+$ is nonvanishing we get $\xi^-=0$ and the 
constraint becomes
\be
\partial_1(\Xi^a\Xi_a-\Xi^{a'}\Xi_{a'})=0.
\ee

Combining it with the  primary constraint $\Xi^2-1=0$ allows to 
obtain
\be
\partial_1(\Xi^a\Xi_a)=\partial_1(\Xi^{a'}\Xi_{a'})=0,
\ee
or equivalently
\be
P^a\Xi_a=P^{a'}\Xi_{a'}=0.\label{tran1}
\ee

It will be useful in the following to 
decompose the variables in Fourier modes
\ba
X^\mu(\sigma)&=&x^\mu+\sum_{n=1}^{\infty}{1 \over \sqrt{4\pi n}}
[(x_n^\mu +\tilde x_n^{\mu\dagger}) e^{-i n\sigma}
 +(x_n^{\mu\dagger} +\tilde x_n^{\mu}) e^{i n\sigma}],\nonumber
 \\
 Q^\mu(\sigma)&=&q^\mu-i\sum_{n=1}^{\infty}\sqrt{n \over{4\pi}}
[(x_n^\mu -\tilde x_n^{\mu\dagger}) e^{-i n\sigma}
 -(x_n^{\mu\dagger} -\tilde x_n^{\mu}) e^{i n\sigma}],\nonumber \\
 \Xi^\mu(\sigma)&=&\xi^\mu+\sum_{n=1}^{\infty}
 {1 \over \sqrt{4\pi n}}
[(\xi_n^\mu +\tilde \xi_n^{\mu\dagger}) e^{-i n\sigma}
 +(\xi_n^{\mu\dagger} +\tilde \xi_n^{\mu}) e^{i n\sigma}], 
 \nonumber
 \\
 P^\mu(\sigma)&=&p^\mu-i\sum_{n=1}^{\infty}\sqrt{n \over{4\pi}}
[(\xi_n^\mu -\tilde \xi_n^{\mu\dagger}) e^{-i n\sigma}
 -(\xi_n^{\mu\dagger} -\tilde \xi_n^{\mu}) e^{i n\sigma}].
 \ea
The Fourier modes satisfy the Poisson Brackets
\be
\{x_n^\mu,\xi^{\nu\dagger}_m\} \ =\ 
\{\tilde x_n^\mu,\tilde \xi^{\nu\dagger}_m\}\ =\ 
\{\xi_n^\mu,x_m^{\nu\dagger}\}\ =\ 
\{\tilde \xi_n^\mu,\tilde x_m^{\nu\dagger}\}\ =\ 
i\delta_{n,m} \eta^{\mu\nu},
\ee
the others being zero.

\subsection{Quantisation}

Let us first recapitulate the first class constraints
that we obtained. They split into left moving ones and right
moving ones. The right moving constraints with their corresponding
conformal weights are
\ba
P^{a'}_R(\sigma)&=&0,\ h=1\label{constra1}\\
P^{-}_R(\sigma)&=&0,\ h=1\label{constra2}\\
\partial(\Xi^a\Xi_a)&=&0,\ h=1\label{constra3}\\
P_R.Q_R&=&0,\ h=2.\label{constra4}
\ea
Since the constraints are first class,
it is possible to quantise with the BRST
prescription (for a review see, e.g., \cite{Henneaux:1985kr}).
 For the quantisation to be possible the BRST charge
must be nilpotent. This in turn implies, as we shall see,
 that the total central
charge,
due to the coordinates and the ghosts, must vanish (see also
\cite{GSW}).
Let the ghosts associated to the constraints
(\ref{constra1}-\ref{constra4}) be denoted respectively by
$c_{a'},c_{-},d$ and $c$ and the corresponding antighosts by
$b^{a'},b^{-},e$ and $b$. All the ghosts except $c$ have conformal
weights $0$ and $c$ has the conformal weight $-1$.
The BRST charge \footnote{recall that if the constraints $G_j$
form a Lie algebra $[G_i,G_j]=C_{ij}{}^{k}G_k$
then the BRST charge is  given by $Q=c^iG_j-{1\over
2}C_{ij}{}^kc^ic^jb_k$.} is given by
\ba
Q=\int d\sigma &\Big[&c(\sigma)T^{(m)} +c_{a'}P^{a'}_R+
c_{-}P^{-}_R+ d\partial(\Xi^a\Xi_a)\nonumber\\
 &+&:c(\sigma)[{1 \over 2}T^{(c)}
+T^{(c_{a'})}+T^{(c_{-})}+T^{(d)}]:\Big],\label{brst}
\ea
where $T^{(m)}$ is the energy-momentum tensor of 
the matter sector and is given by $T^{(m)}=:P_R.Q_R:/2$
with the commutation relations
\be
[T^{(m)}(\sigma),T^{(m)}(\sigma')]=
-i(T^{(m)}(\sigma)+T^{(m)}(\sigma'))
\delta'(\sigma-\sigma')-i{c^m \over 24}\delta'''(\sigma-\sigma'),
\label{commu}\ee
$c^m$ being the central charge of the right moving matter sector
given by $2D$; $T^{(c)}$ is the energy-momentum tensor of the
ghost system $c$ and $b$ and so on for the other terms in 
(\ref{brst}). The energy-momentum tensor of 
the weight $h$ ghost \footnote{Here $h$ is the conformal weight of
the antighosts $b_h$ which is the same as that
of the corresponding constraint; we use the conventions 
$\{c_h(\sigma),b_h(\sigma')\}=\delta(\sigma-\sigma')$.}  system is
given by
\be
T_h=i:[h\partial(c_h
b_h)-c_h\partial b_h]:,\label{gho}
\ee
which satisfies commutation relations analogous to
(\ref{commu})  with a  central charge given by 
\cite{Friedan:1985ge,GSW} $c_h=1-3(2h-1)^2$. 
In the determination of the second line of
(\ref{brst}) we have also used the commutation relations of the
matter energy-momentum tensor with the other constraints:
if we denote generically one of the constraints (\ref{constra1}-
\ref{constra3}) by $K$ then we have
\be
\left[T^{(m)}(\sigma'),K(\sigma)\right]=iK'(\sigma)
\delta(\sigma-\sigma')
+i K(\sigma)\delta'(\sigma-\sigma').
\ee
The crucial condition in the BRST quantisation is the nilpotency
of $Q$, the calculation of $Q^2$, using (\ref{brst}), (\ref{commu})
and (\ref{gho}), gives
\be
Q^2=i{c_T \over 48}\int d\sigma\ (\partial^3c(\sigma))\ c(\sigma),
\ee
where $c_T$ is the total central charge.
Thus the nilpotency of $Q$ requires  
that the total central charge 
is zero. A similar conclusion is of course  true for the left
sector.
The total central charge due to the
ghosts of the right sector is  given by
\be
c_{g\ R}=-26-2(D-N).
\ee
Similarly the total central charge due to the ghosts of the left
movers is
\be
c_{g\ L}=-26-2(N+2).
\ee
Equality of the two central charges gives $N$
\be
N={{D-2}\over 2}.
\ee
The matter contribution to the central charge is $2D$.
The vanishing of the total central charge gives $N$ and $D$
\be
N=13, \ D=28.
\ee

In order to determine the physical states it is more convenient
to fix the gauge invariance associated with the first class
constraints.  As is well known to each first class constraint
is associated a gauge invariance, and complete gauge
fixing amounts to adding constraints such that the whole system be
second class.
The unconstrainted phase space has $4D$ degrees of freedom,
$X^\mu,P^\mu,\Xi^\mu$ and $Q^\mu$. 
We have $D+4$ first class constraints, gauge fixing
introduces $D+4$ further constraints and we are
left with a total of $2(D-4)$ degrees of freedom. These correspond
to $D-4=24$ coordinates, exactly the same number as
the physical coordinates of the bosonic string.
This fact renders the model a serious candidate for the
description of the tensionless limit of the usual bosonic string.
Although the spacetime dimensions are different, the number of degrees
of freedom is the same. This deserves further study.

It is straightforward to verify that it is possible to
 choose the following gauge fixing 
constraints
\ba
Q^{a'}+\partial_1X^{a'}=q^{a'},\quad Q^{a}-\partial_1X^{a}=q^{a},\\
Q^+=0,\quad X^+=0\\
P^aQ_a=0,\quad P^{a'}Q_{a'}=0,\label{tran2}\\
P^{+}=p^+,\quad \Xi^{+}=\xi^+.
\ea
Here $q^{i}, \xi^+$ and $p^+$ are $\sigma$ independent.
Using these constraints in the Virasoro constraints
gives $Q^-$ and $\partial_1X^-$ as
\be
p^+Q^-+P^aq_a+P^{a'}q_{a'}=a,\quad
p^+\partial_1X^-+P^aq_a-P^{a'}q_{a'}=0,
\ee
where we allowed for a normal ordering constant.
If we take the zero mode part of the first equation we discover
that $p^+q^-=p.q=a$, that is physical states are all massless
and belong  the continuous spin
representation of the Poincar\'e group. 

As we noted in Section 4, the $(D,D)$ signature of the kinetic
terms in the string action can potentially lead to negative norm
states. After gauge fixing, the only physical variables
are $\xi_n^a,\xi^{a\dagger}_n, x_n^{a},x^{a\dagger}_n$
and $\tilde \xi_n^{a'},\tilde\xi^{a'\dagger}_n, \tilde x_n^{a'}, 
\tilde x^{a'\dagger}_n$ as well as the zero modes.
The other variables are eliminated by the second class
constraints. The Dirac brackets \footnote{For simplicity we did
not take into account
the constraints (\ref{tran1}) and their ``conjugate"
(\ref{tran2})} are 
transformed to commutators by the quantisation prescription:
$[\xi^{a}_n,x_m^{b\dagger}]=\delta^{ab}\delta_{mn}$ and so on.
One is thus tempted to interpret the daggered operators as
creation operators and the undaggered ones annihilation operators
and define the vacuum to be the state annihilated by 
all the annihilation operators.
This, however, does not lead to a positive definite Hilbert space.
Consider for example the state obtained by acting with 
$x^{a\dagger}_n-\xi^{a\dagger}_n$ on the vacuum.
It has a negative norm (if the vacuum has a positive norm as it
should). There is a way out of this problem which maintains all
the desired symmetries. Define the vacuum as the state 
which verifies
\ba
(x^{a\dagger}_n-\xi^{a\dagger}_n)|0>&=&0,\
(x^{a}_n+\xi^{a}_n)|0>=0\\
(\tilde x^{a'\dagger}_n-\tilde \xi^{a'\dagger}_n)|0>&=&0,\
(\tilde x^{a'}_n+\tilde \xi^{a'}_n)|0>=0.
\ea
This leads to a positive definite Hilbert space.
In fact define
\be
z^a_n={1\over \sqrt{2}}(x^{a\dagger}_n-\xi^{a\dagger}_n),\
y^a_n={1\over \sqrt{2}}(x^{a}_n+\xi^{a}_n),
\ee
and similarily for tilded operators, then
we have
\ba
\left[z^a_n, z^{b\dagger}_m\right]&=&\delta^{ab}\delta_{mn}=
\left[y^{a}_n,y^{b\dagger}_m\right],\\
\left[z^a_n, z^{b}_m\right]&=&0=
\left[y^{a}_n,y^{b}_m\right],
\ea
the commutators between the $y's$ and the $z'$ are zero.
The Fock space obtained by using the $z^{a}_n$
and $y^{a}_n$ as annihilation operators is thus manifestly
positive definite.
Notice that the constraints were crucial to render this possible:
they allowed the elimination of the two time like coordinates
$X^0$ and $\Xi^0$.

\section{Discussion}

The quantization of the Polyakov string action can be done either 
in the BRST formalism or in the light cone gauge.
In the first case, one has manifestly Lorentz covariant
first class constraints and the critical dimension appears as the
condition of nilpotency of the BRST charge.
In the second quantisation scheme, one adds gauge fixing
conditions
and solves the constraints at the price of breaking manifest 
Lorentz
invariance. The critical dimension appears as the condition 
of Lorentz invariance at the quantum level.
In the present case,
we have a mixture of the two above formalisms;
the transformation of the bilocal constraints
to local ones was done at the expense of breaking manifest Lorentz
covariance. 
We ended with a Lie algebra of first class constraints
and the nilpotency condition of
the BRST charge gave us the critical dimension of 28.
This condition is equivalent to the vanishing of the total central
charge, the anomaly of the Weyl symmetry.
What we have thus shown is that a {\it necessary}
 condition for the
consistency
of the theory is that the space-time dimension be 28.
 In order to fully prove the consistency of
 this BRST quantisation one has in addition to
 prove that in this critical dimension Lorentz
covariance is maintained at the quantum level \footnote{
 Progress on this issue was obtained recently
 in {\it J. Mourad, Continuous spin particles from a string theory, 
 hep-th 0504118}, where the physical
 spectrum in
 the BRST formalism was shown to be Lorentz covariant}.

 It is also of interest to study the spectrum and its
 relation with the tensionless limit of the bosonic string.
The study of interactions, along the lines started 
in \cite{Savvidy:2004bb}
 is also of great interest. We hope to come back soon
 to some of these issues.  



\begin{thebibliography}{999}

\bibitem{Wigner:1939cj}
E.~P.~Wigner,
Annals Math.\  {\bf 40} (1939) 149
[Nucl.\ Phys.\ Proc.\ Suppl.\  {\bf 6} (1989) 9].

\bibitem{wi}
E.P. Wigner, Z. Physik {\bf 124} (1947) 665.


\bibitem{Bargmann:1948ck}
V.~Bargmann and E.~P.~Wigner,
Proc.\ Nat.\ Acad.\ Sci.\  {\bf 34}, 211 (1948).

\bibitem{chak}J. Yngvason, Commun. Math. Phys. 18 (1970) 195;
A. Chakrabarty, J. Math. Phys. 12, 1813 (1971); G.J. Iverson and G.
Mack, Annals of Physics 64 (1971) 211.

\bibitem{Abbott:1976bb}
L.~F.~Abbott,
Phys.\ Rev.\ D {\bf 13}, 2291 (1976);
K.~Hirata,
Prog.\ Theor.\ Phys.\  {\bf 58} (1977) 652.

\bibitem{Mund:2004sy}
J.~Mund, B.~Schroer and J.~Yngvason,
Phys.\ Lett.\ B {\bf 596} (2004) 156
[arXiv:math-ph/0402043].

\bibitem{Zoller:1991hs}
D.~Zoller,
Class.\ Quant.\ Grav.\  {\bf 11}, 1423 (1994).



\bibitem{Brink:2002zx}
L.~Brink, A.~M.~Khan, P.~Ramond and X.~z.~Xiong,
J.\ Math.\ Phys.\  {\bf 43} (2002) 6279
[arXiv:hep-th/0205145].

\bibitem{Zoller:1990jv}
M.~S.~Plyushchay,
Mod.\ Phys.\ Lett.\ A {\bf 4} (1989) 837;
D.~Zoller,
Phys.\ Rev.\ Lett.\  {\bf 65}, 2236 (1990).

\bibitem{Savvidy:2003dv}
G.~K.~Savvidy,
Phys.\ Lett.\ B {\bf 552} (2003) 72.

\bibitem{Savvidy:2003fx}
G.~K.~Savvidy,
Int.\ J.\ Mod.\ Phys.\ A {\bf 19} (2004) 3171
[arXiv:hep-th/0310085].

\bibitem{Antoniadis:2004uh}
I.~Antoniadis and G.~Savvidy,
arXiv:hep-th/0402077.

\bibitem{Savvidy:2004bb}
G.~Savvidy,
arXiv:hep-th/0409047.

\bibitem{Becchi:1975nq}
C.~Becchi, A.~Rouet and R.~Stora,
Annals Phys.\  {\bf 98} (1976) 287;
I.V. Tyutin, Lebedev Institute preprint N39 (1975).
\bibitem{Henneaux:1985kr}
M.~Henneaux,
Phys.\ Rept.\  {\bf 126} (1985) 1.


\bibitem{Schild:1976vq}
A.~Schild,
Phys.\ Rev.\ D {\bf 16} (1977) 1722.

\bibitem{Karlhede:1986wb}
A.~Karlhede and U.~Lindstrom,
Class.\ Quant.\ Grav.\  {\bf 3}, L73 (1986);
F.~Lizzi, B.~Rai, G.~Sparano and A.~Srivastava,
Phys.\ Lett.\ B {\bf 182} (1986) 326;
J.~Isberg, U.~Lindstrom and B.~Sundborg,
Phys.\ Lett.\ B {\bf 293}, 321 (1992)
[arXiv:hep-th/9207005];
U.~Lindstrom, B.~Sundborg and G.~Theodoridis,
Phys.\ Lett.\ B {\bf 253}, 319 (1991);
S.~Hassani, U.~Lindstrom and R.~von Unge,
Class.\ Quant.\ Grav.\  {\bf 11}, L79 (1994);
JHEP {\bf 0201}, 034 (2002)
[arXiv:hep-th/0112206].

\bibitem{Sundborg:2000wp}
B.~Sundborg,
Nucl.\ Phys.\ Proc.\ Suppl.\  {\bf 102} (2001) 113
[arXiv:hep-th/0103247];
U.~Lindstrom and M.~Zabzine,
Phys.\ Lett.\ B {\bf 584} (2004) 178
[arXiv:hep-th/0305098];
G.~Bonelli,
Nucl.\ Phys.\ B {\bf 669} (2003) 159
[arXiv:hep-th/0305155];
A.~Sagnotti and M.~Tsulaia,
Nucl.\ Phys.\ B {\bf 682} (2004) 83
[arXiv:hep-th/0311257].

\bibitem{Dirac:1950pj}
P.~A.~M.~Dirac,
Can.\ J.\ Math.\  {\bf 2} (1950) 129;
for a review see 
K.~Sundermeyer,
Lect.\ Notes Phys.\  {\bf 169} (1982) 1.



\bibitem{Bouatta:2004kk}
N.~Bouatta, G.~Compere and A.~Sagnotti,
arXiv:hep-th/0409068.

\bibitem{Friedan:1985ge}
D.~Friedan, E.~J.~Martinec and S.~H.~Shenker,
Nucl.\ Phys.\ B {\bf 271} (1986) 93.

\bibitem{GSW} M. Green, J. Schwarz and E. Witten, 
{Superstring theory}, Vol. 1, Cambridge university press (1987),
chap. 3;
J. Polchinski, {String theory}, Vol. 1, Cambridge university
press (1998).
\end{thebibliography}
\end{document}